\documentclass[aps,prb,twocolumn,superscriptaddress,showpacs]{revtex4-1}
\usepackage{graphicx,amsmath,color,placeins}

\begin{document}
	\title{Spin-Triplet Topological Excitonic Insulators in Two-dimensional Materials}
	\author{Huaiyuan Yang}
	\affiliation{State Key Laboratory for Artificial Microstructure and Mesoscopic Physics, Frontier Science Center for Nano-optoelectronics and School of Physics, Peking University, Beijing 100871, P. R. China}
	\author{Jiaxi Zeng}
	\affiliation{State Key Laboratory for Artificial Microstructure and Mesoscopic Physics, Frontier Science Center for Nano-optoelectronics and School of Physics, Peking University, Beijing 100871, P. R. China}
	\author{Yuelin Shao}
	\email{ylshao@iphy.ac.cn}
	\affiliation{Beijing National Laboratory for Condensed Matter Physics and Institute of Physics, Chinese Academy of Sciences, Beijing 100190, P. R. China}
	\affiliation{School of Physical Sciences, University of Chinese Academy of Sciences, Beijing 100049, P. R. China}
	\author{Yuanfeng Xu}
	\affiliation{Center for Correlated Matter, School of Physics, Zhejiang University, No.866, Yuhangtang Road, Xihu District, Hangzhou, Zhejiang Province, PRC}
	\author{Xi Dai}
	\affiliation{Department of Physics, Hong Kong University of Science and Technology, Clear Water Bay, Kowloon 999077, Hong Kong}
	\author{Xin-Zheng Li}
	\email{xzli@pku.edu.cn}
	\affiliation{State Key Laboratory for Artificial Microstructure and Mesoscopic Physics, Frontier Science Center for Nano-optoelectronics and School of Physics, Peking University, Beijing 100871, P. R. China}
	\affiliation{Interdisciplinary Institute of Light-Element Quantum Materials, Research Center for Light-Element Advanced Materials, and Collaborative Innovation Center of Quantum Matter, Peking University, Beijing 100871, People's Republic of China}
	\affiliation{Peking University Yangtze Delta Institute of Optoelectronics, Nantong, Jiangsu 226010, People's Republic of China}
	\date{\today}

	\begin{abstract}
		Quantum spin-hall insulator (QSHI) processes nontrivial topology.
		We notice that the electronic structures of some particular QSHIs are favorable for realization of excitonic insulators (EIs).
		Using first-principles many-body perturbation theory ($GW$+BSE) and $k \cdot p$ model, we show that high-temperature ($T$) topological EIs with unlike spin can exist in
		such QSHIs with non-vanishing band gaps, e.g. 2D AsO and $\text{Mo}_2\text{Ti}\text{C}_2\text{O}_2$.
		Spin-triplet type EI phase induced by strong electron-hole interaction preserves time-reversal symmetry and the topological characteristics.
		A novel optical selection rule exists, upon going through the phase transition from the normal QSHIs to the topological EIs, absorption
		spectroscopy shows pronounced $T$-dependent changes, providing guidance for future experimental detections.
		The demonstrated coupling between EIs and topology also means that rich physics exists in such materials which retain such interdisciplinary features.

	\end{abstract}
	\maketitle
	\clearpage

	The quantum spin hall insulator (QSHI) is a kind of two-dimensional (2D) state of matter, which is topologically different from normal insulators
	due to band inversion ~\cite{PhysRevLett.95.226801, PhysRevLett.95.146802, PhysRevLett.100.236601, doi:10.1126/science.1133734}.
	They have insulating bulk gaps and gapless edge states, which support dissipationless helical transport at the edges.
	A $\text{Z}_2$ topological index is needed to identify their nontrivial topology.
	Up to now, there have been plenty of theoretical investigations and experimental realizations on these QSHIs~\cite{doi:10.1021/acs.jpclett.7b00222, https://doi.org/10.1002/adma.202008029}.
	Similar to the QSHI, excitonic insulator (EI) is another quantum state of matter which has received intensive research attention in
	recent years~\cite{du2017evidence,wang2019evidence,wakisaka2009excitonic, lu2017zero, mor2017ultrafast, jiang2018realizing, jiang2019half, jiang2020spin, Varsano2020367, Yang_2022}.
	This concept was proposed by Keldysh and Kohn in the 1960s~\cite{keldysh1965possible, kohn1967excitonic, jerome1967excitonic},
	where bound excitons composited by electron-hole pairs form and condensate as the exciton binding energy $E_{\text{b}}$ exceeds the band gap $E_{\text{g}}$.
	In the EIs, detectable quasiparticle gap-opening resembles the superconductors, but no charge transport exists~\cite{RevModPhys.40.755}.
	This seriously hinders their experimental detection by transport measurement.
	Over the last half century, massive efforts have been made to realize the EI in real materials, and much progress have been achieved.
	There have been several experimental indications for EIs in materials with lattice instability such as $1T$-TiSe$_2$~\cite{cercellier2007evidence, kogar2017signatures} and Ta$_2$NiSe$_5$~\cite{wakisaka2009excitonic, lu2017zero, mor2017ultrafast}, in artificial heterostructures such as InAs/GaSb quantum well~\cite{du2017evidence} and $\text{MoSe}_{2}/\text{WSe}_{2}$ bilayers~\cite{wang2019evidence}, and in stable monolayer $\text{WTe}_2$~\cite{sun2022evidence, jia2022evidence}.
	Nevertheless, it is fair to say that a systematic strategy to identify the EI phase is still absent.
	From the theoretic point of view, realization of the EI phase requires reduced screening of the Coulomb interactions, which is drastic in two-dimensional systems.
	However, 2D semiconductors with large $E_{\text{b}}$ tends to have large $E_{\text{g}}$, as both of them are inversely proportional to the magnitude of
	screening~\cite{jiang2017scaling}.
	To break this synergy, one strategy is to seek for dipole forbidden transitions near the band edges~\cite{jiang2018realizing, jiang2019half, jiang2020spin}.
	Interestingly, we note that this condition can be satisfied in some particular inversion-symmetric QSHIs with same-parity band-edge states
	~\cite{doi:10.1021/acs.nanolett.6b03118, doi:10.1063/1.4983781}.
	Normally, the parity of the conduction band minimum (CBM) and the valence band maximum (VBM) is different in a QSHI.
	However, a scenario may exist when the band inversion does not happen between the CBM and VBM, but between a far-away band and the CBM+VBM together.
	Spin-orbital coupling (SOC) then opens a gap between the CBM and VBM.
	In this way, the band-edge states possess the same parity so that the dipole transitions are forbidden, and the EI phase may be achieved.
	So far, not much efforts have been made on realizing EIs in such topological materials~\cite{PhysRevLett.112.146405, PhysRevLett.112.176403, PhysRevLett.120.186802, Varsano2020367}.
	It is reasonable for us to expect robust EIs protected by symmetry and some new physics in this interdisciplinary area.
	In this letter, we demonstrate that these desired excitonic instability with topological features can exist in such QSHI materials like 2D arsenene oxide (AsO)
	and $\text{Mo}_2\text{Ti}\text{C}_2\text{O}_2$.
	First-principles $GW$ approximation and Bathe-Salpeter equation (BSE) results show that the exciton binding energies ($E_{\text{b}}$) are larger
	than the $GW$ band gaps ($E_{\text{g}}$).
	Then, using $k \cdot p$ models derived from the $GW$ bandstructures, we analyze the properties of the excitons and their condensation, i.e. the EIs, in detail.
	An optical selection rule, qualitatively different from excitons in conventional semiconductors, is identified.
	The order parameters of the EI phase, obtained from self-consistent calculations by taking the electron-hole interactions into account, show that the final
	exciton condensation is of the spin-triplet type, which can coexist with nontrivial topology.
	This peculiar spin-triplet topological excitonic insulator (TEI) phase can exist at very high temperatures ($T$s) in realistic
	screening environments, as unveiled by absorption spectra, where pronounced $T$-dependent changes exist.
	These drastic changes of the bandstructures and the absorption spectra upon transition from the QSHI to the TEI phase provide experimental fingerprints
	for detection of such TEI phases in future experiments.
	The coupling between EIs and topology demonstrated also means that rich physics can exist in materials retaining such interdisciplinary features.
	In the main manuscript, we show results of 2D AsO.
	2D  $\text{Mo}_2\text{Ti}\text{C}_2\text{O}_2$ shares very similar low-energy $k \cdot p$ model and consequently very similar excitonic and topological properties,
	as demonstrated in the Supplementary Material (SM) together with the details of the first-principles calculations.
	Fig. 1(a) shows the lattice structure of AsO, which is similar to hexagonal 2D decorated stanene with inversion-symmetry~\cite{doi:10.1063/1.4983781, PhysRevLett.111.136804}.
	Oxygen plays the role of chemical functional group.
	The band inversion happens between the $p_{z}$ orbital with negative parity and the $p_{x,y}$ with positive one~\cite{doi:10.1063/1.4983781}.
	Unlike many topological insulators, the SOC does not induce the band inversion, but contributes to open a band gap between the
	$p_{x,y}$ orbitals near the Fermi level~(Fig.~1(b))~\cite{doi:10.1063/1.4983781}.
	The $GW$ bandstructure does not change significantly from the PBE one except that the band gap value increases from 93 meV to 165 meV.
	After solving the BSE, we obtain a large exciton binding energy, $\sim$300 meV, as the band-edge states have the same parity~\cite{jiang2018realizing}.
	This $E_{\text{b}}$ value exceeds the $GW$ band gap~(Fig.~1(c)), indicating excitonic instability.
	To investigate the excitonic properties in more details, we build a 4-band low-energy $k \cdot p$ model using the four states near Fermi level (in the order of $\left\{\left|p_x+ip_y, \uparrow\right\rangle,\left|p_x+ip_y, \downarrow\right\rangle,\left|p_x-ip_y, \uparrow\right\rangle,\left|p_x-ip_y, \downarrow\right\rangle\right\}$).
	Including SOC, the Hamiltonian reads:
	\begin{equation}\label{kp}
		H_0=
		\left(\begin{array}{cccc}
			h_{0}(\boldsymbol{k})+\lambda & 0 & c_2 k_-^2 & 0 \\
			0 & h_{0}(\boldsymbol{k})-\lambda & 0 & c_2 k_-^2 \\
			c_2 k_+^2 & 0 & h_{0}(\boldsymbol{k})-\lambda & 0 \\
			0 &c_2 k_+^2 & 0 & h_{0}(\boldsymbol{k})+\lambda
		\end{array}\right).
	\end{equation}
	Here $h_{0}(\boldsymbol{k})=\epsilon_p+c_1\left(k_x^2+k_y^2\right), k_{\pm}=k_x \pm iky$, and $\lambda$ is the SOC strength.
	It can be written in a more compact form as
	\begin{equation}
		\begin{gathered}
			H_0=h_{0}(\boldsymbol{k})\tau_0s_0 + c_2 (k_x^2-k_y^2) \tau_x s_0 \\
			+ 2c_2 k_xk_y \tau_y s_0+\lambda \tau_z s_z
		\end{gathered}
	\end{equation}
	where $\tau_i$ and $s_i (i=0, x, y, z)$ are the Pauli matrices acting on the $p_x \pm ip_y$ orbital space and the up/down spin space.
	This Hamiltonian is inversion symmetric ($I$), i.e. $H_0(k)=H_0(-k)$, and is invariant under time-reversal symmetry $\mathcal{T}=\tau_x s_y K$.
	The parameters $\epsilon_p, c_1, c_2$ and $\lambda$ are fitted to reproduce the $GW$ bandstructure near the $\Gamma$ point.
	From Fig. 1(c), we see that our chosen values can provide accurate matching between the model and the first-principles results.
	As $H_0$ is spin-diagonal, we can explicitly calculate the Chern number for both spin sectors by integrating the Berry curvature using
	$C_{\uparrow / \downarrow}=\frac{1}{2 \pi} \iint \mathcal{F}_{\uparrow / \downarrow} d k_x d k_y$.
	$\mathcal{F}_{\uparrow}$ is shown in Fig.~1(d), which is nonzero around $\Gamma$, giving $C_{\uparrow}=1$.
	The down-spin Chern number can be similarly obtained, which is $-1$.
	Therefore the spin Chern number is $C_s=\frac{1}{2}(C_{\uparrow}-C_{\downarrow})=1$, signaling the nontrivial topology of the system, and in alignment with the analysis of the bandstructure.

	\begin{figure}[h]
		\includegraphics[width=1.0\linewidth]{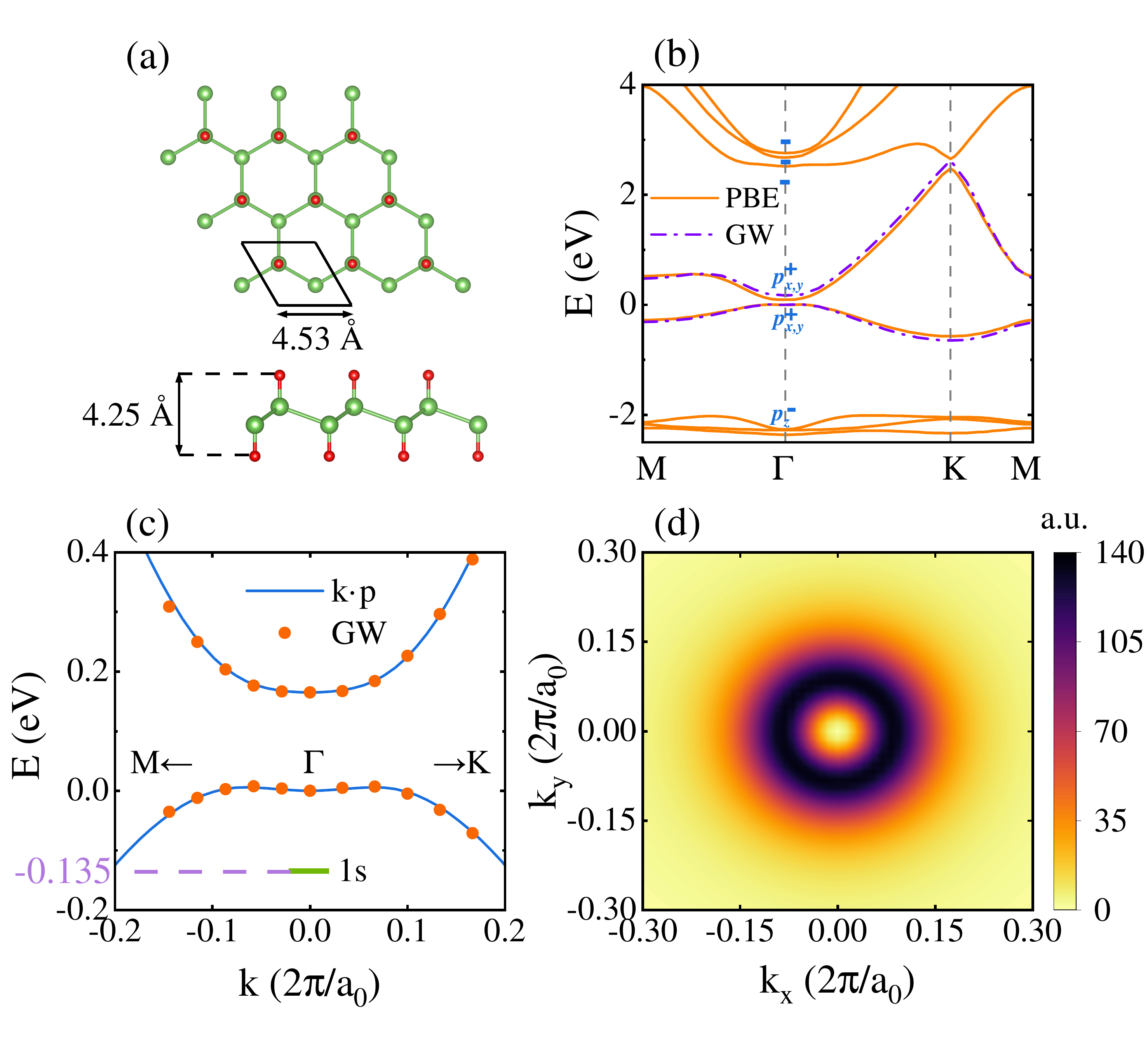}
		\caption{ \label{fig1}
			(a) Top (top) and side (bottom) views of the atomic structures of AsO.
			Green and red balls denote the As and O atoms respectively.
			(b) The PBE (orange solid lines) and $GW$ (purple dot-dashes lines) bandstructures of AsO.
			The blue symbols mark the parity of the bands near the Fermi level.
			(c) Low-energy $k \cdot p$ model (blue solid lines) fitted from GW results (orange dots).
			The 1s exciton energy level is marked by the short green line.
			(d) Berry curvature for spin-up sector calculated from the $k \cdot p$ model.
		}
	\end{figure}

	On top of this single-particle picture of this low-energy $k \cdot p$ model, we performed BSE calculations to explore the excitonic properties.
	The BSE reads~\cite{PhysRevB.62.4927}
	\begin{equation}\label{bse}
		(E_S-\epsilon_{c}(\boldsymbol{k})+\epsilon_{v}(\boldsymbol{k})) A^S_{c v}(\boldsymbol{k})= \sum_{c^{\prime}, v^{\prime}, \boldsymbol{k}^{\prime}} \mathcal{K}_{c v \boldsymbol{k}, c^{\prime} v^{\prime} \boldsymbol{k}^{\prime}} A^S_{c^{\prime}, v^{\prime}}(\boldsymbol{k}^{\prime}),
	\end{equation}
	where $\epsilon_{c}(\boldsymbol{k})$ ($\epsilon_{v}(\boldsymbol{k})$) is the quasiparticle energy of the conduction (valence) band,
	and $|S\rangle=\sum_{c v \boldsymbol{k}} A^S_{c v}(\boldsymbol{k}) \hat{c}^{\dagger}_{c \boldsymbol{k}} \hat{c}_{v \boldsymbol{k}} |\text{GS}\rangle$ is the exciton eigenstate.
	The kernel is consisted of direct and exchange part~\cite{PhysRevB.62.4927, PhysRevLett.119.127403},
	$\mathcal{K}_{c v \boldsymbol{k}, c^{\prime} v^{\prime} \boldsymbol{k}^{\prime}}=\mathcal{K}_{c v \boldsymbol{k}, c^{\prime} v^{\prime} \boldsymbol{k}^{\prime}}^{\mathrm{d}}+\mathcal{K}_{c v \boldsymbol{k}, c^{\prime} v^{\prime} \boldsymbol{k}^{\prime}}^{\mathrm{x}}$,
	where
	\begin{equation}\label{kernel}
		\begin{aligned}
			&\mathcal{K}_{c v \boldsymbol{k}, c^{\prime} v^{\prime} \boldsymbol{k}^{\prime}}^{\mathrm{d}}=-\frac{W(\boldsymbol{k}-\boldsymbol{k}^{\prime}) f_{c c^{\prime}}(\boldsymbol{k}, \boldsymbol{k}^{\prime}) f_{v^{\prime} v}(\boldsymbol{k}^{\prime}, \boldsymbol{k})}{\Omega}, \\
			&\mathcal{K}_{c v \boldsymbol{k}, c^{\prime} v^{\prime} \boldsymbol{k}^{\prime}}^{\mathrm{x}}=-\frac{V f_{c v}(\boldsymbol{k}, \boldsymbol{k}) f_{v^{\prime} c^{\prime}}(\boldsymbol{k}^{\prime}, \boldsymbol{k}^{\prime})}{\Omega}.
		\end{aligned}
	\end{equation}
	Here $\Omega$ is the system area, $V$ is the bare Coulomb interaction, and $W(\boldsymbol{q})$ is the screened Coulomb interaction, which equals $2 \pi/\left[ |q|\left(1+2 \pi |q| \alpha_{2 \mathrm{D}}\right)\right]$.
	$\alpha_{2 \mathrm{D}}$ is the 2D polarizability obtained from first-principles random-phase approximation (RPA) calculation.
	The form factors $f_{n n^{\prime}}(\boldsymbol{k}, \boldsymbol{k}^{\prime})=\left[\eta_{n \boldsymbol{k}}\right]^{\dagger} \eta_{n^{\prime} \boldsymbol{k}^{\prime}}$ are calculated from the single-particle eigenstates $\eta_{n \boldsymbol{k}}$.
	Therefore $f_{c v}(\boldsymbol{k}, \boldsymbol{k})$ equals 0 due to orthogonality of the eigenstates, resulting in $\mathcal{K}_{c v \boldsymbol{k}, c^{\prime} v^{\prime} \boldsymbol{k}^{\prime}}^{\mathrm{x}}=0$.
	This means that we can consider the like-spin ($\uparrow \uparrow$ and $\downarrow \downarrow$) and unlike-spin ($\uparrow \downarrow$ and $\downarrow \uparrow$) excitons separately~\cite{PhysRevB.62.4927}.
	Here we label the spin of the occupied state by the spin of the electron that originally occupies it.
	The binding energy obtained from this model BSE calculation is $\sim$300 meV, matching the first-principles $GW$-BSE result.
	The energy of the unlike-spin exciton is several meV lower than the like-spin excitons, also in alignment with the first-principles one.
	These facts mean that the model calculation can capture the main physics we are concerned about.
	We choose the gauge by requiring the envelope function $A(\boldsymbol{k})$ of the lowest-energy 1s-like exciton as real all over the $k$-space.
	In so doing, the envelope functions of the excitons are hydrogenlike characterized by a series of quantum numbers $m$s.
	To be specific, the envelope function for each excitonic state has the following form: $A^{m}(\boldsymbol{k})=\tilde{A}^{m}(|\boldsymbol{k}|) e^{im\phi_{\boldsymbol{k}}}$, where $\phi_{\boldsymbol{k}}$ is the angle between $\boldsymbol{k}$ and $k_x$ axis.
	Taking the lowest three $\uparrow \uparrow$ excitons as examples (Fig.~2(a,b,c)), we see that the absolute value of the wave functions are of the 1s or 2p type and localized
	around the $\Gamma$ point.
	They all exhibit Wannier characteristics, and their phase winding number $m$ equals 0 and $\mp1$ respectively.
	Now we look at the optical properties in this model.
	The oscillator strength of the right- and left-handed photon modes ($\sigma_\pm$) reads~\cite{PhysRevLett.120.087402, PhysRevLett.120.077401}:
	\begin{equation}\label{optical}
		\mathrm{I}_{\sigma_{\pm}}^S=2\left|\Sigma_{\boldsymbol{k}} A^S(\boldsymbol{k}) \cdot p_{\boldsymbol{k}, \mp} \right|^2/E_S,
	\end{equation}
	where $p_{\boldsymbol{k}, \pm} = \left\langle v \boldsymbol{k}\left| \hat{p}_x \right| c \boldsymbol{k}\right\rangle \pm  \left\langle v \boldsymbol{k}\left|\hat{p}_y \right| c \boldsymbol{k}\right\rangle$.
	The complex quantity $p_{\boldsymbol{k}, \pm}$ in spin-up sector is shown by shades of color which represents the magnitude, and arrows pointing along the phase angle in Fig.~2(d,e).
	The absolute values of $p_{\boldsymbol{k}, +}$ and $p_{\boldsymbol{k}, -}$ are both isotropic while the latter is much larger in magnitude than the former.
	Their phase winding patterns share the similar feature with the exciton wavefunctions, i.e. has a cylindrical angular phase dependence in the form of $\sim e^{il_{+/-}\phi_{\boldsymbol{k}}}$, and $l_{+}=+3, l_{-}=+1$.
	As a result, Eq.~\ref{optical} can be written as $\mathrm{I}_{\sigma_{\pm}}^S=2\left|\Sigma_{\boldsymbol{k}} {f}(|\boldsymbol{k}|) e^{i(m+l_{+/-})\phi_{\boldsymbol{k}}} \right|^2/E_S$, where ${f}(|\boldsymbol{k}|)$ is the radial part in the integrand.
	Therefore $\mathrm{I}_{\sigma_{\pm}}^S$ is nonzero when $m=-l_{-}=-1$ or $m=-l_{+}=-3$.
	After numerical calculations, it is found that $\mathrm{I}_{\sigma_{+}}^{m=-3}$ is much smaller than $\mathrm{I}_{\sigma_{-}}^{m=-1}$, and we can conclude that the 2p-like exciton with ${m=-1}$ is the only bright exciton among all $\uparrow \uparrow$ excitons, coupling with the $\sigma_+$ photon modes.
	Due to the time-reversal symmetry, for $\downarrow \downarrow$ excitons, $m=+1$ exciton is the only bright one which couples with $\sigma_-$ photon modes, and therefore a spin circular dichroism exists.
	These conclusions donot change when we consider the $C_3$ symmetry of the system, as the magnitude of higher Fourier component generally decays to zero fast~\cite{PhysRevLett.120.087402, PhysRevLett.120.077401}.
	This is also supported by the first-principles BSE results.
	The exotic optical selection rule is distinct from the normal semiconductors, but in alignment with the optical selction rule for chiral fermions with $|w|=2$ proposed in Ref. \onlinecite{PhysRevLett.120.077401}, and is demonstrated schematically in Fig.~2(f).
	This rule means that the 1s-like excitons with the lowest energy are dark (Fig.~2(a)), indicating larger lifetimes than those of the bright exciton which is favorable
	for condensation.

	\begin{figure}[h]
		\includegraphics[width=1.0\linewidth]{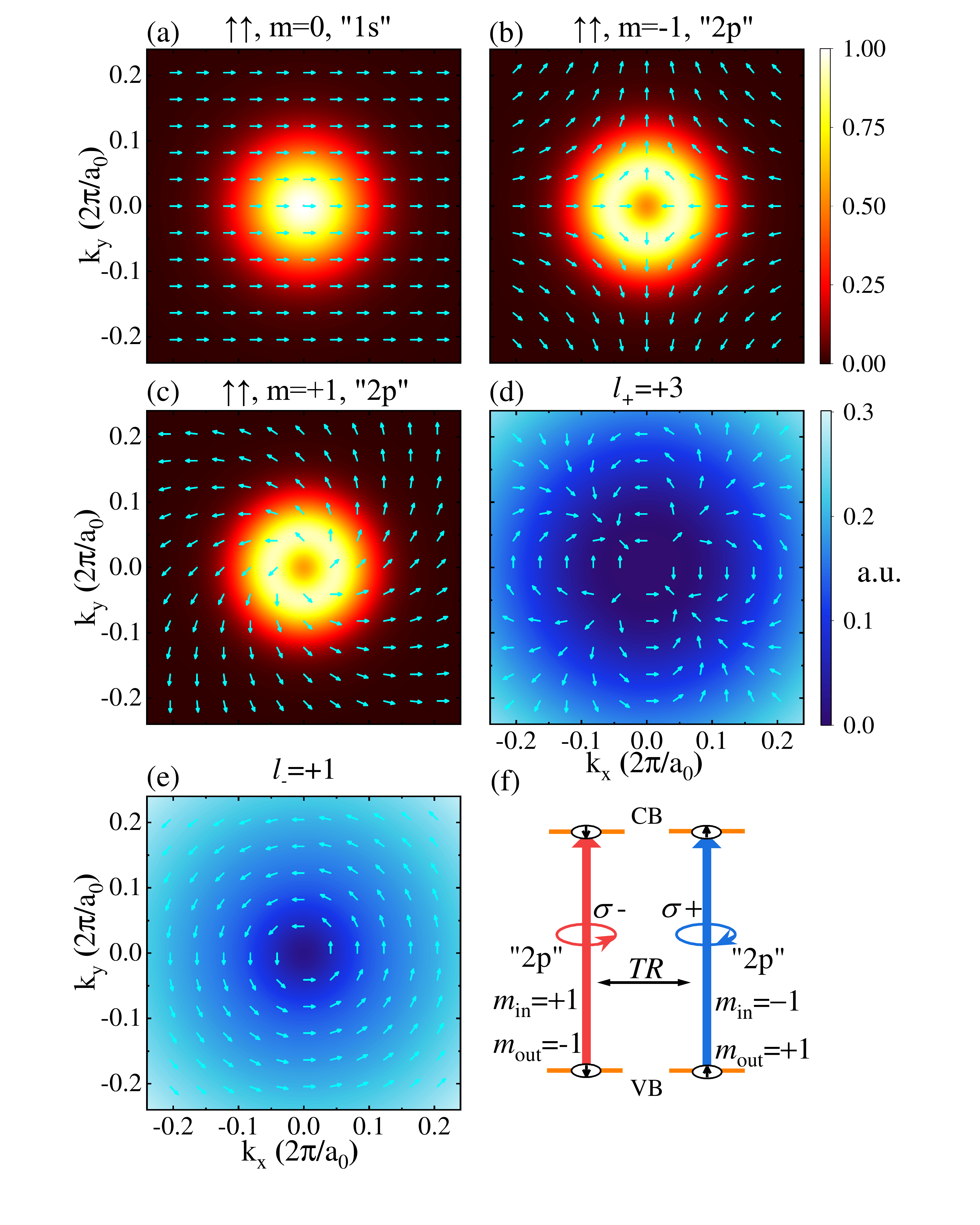}
		\caption{ \label{fig2}
			(a)-(c) Absolute values (colors) and the phase angles (arrows) of the lowest three $\uparrow \uparrow$ exciton wavefunction in $k$-space.
			The largest  absolute value of each exciton wavefunction is set to unity.
			Absolute value of the interband optical transition matrix element and its phase winding pattern of the spin-up sector for light with (d) left circular polarization $p_{\boldsymbol{k}, +}$ and (e) right circular polarization $p_{\boldsymbol{k}, -}$.
			The direction of the arrow denotes the phase angle and the colormap indicates the magnitude.
			The effective phase winding number is determined by inspecting how much the angle which the arrow points at increases when the phase angle of $\mathbf{k}$ increases $\pi/2$ in each quarter.
			(f) Schematics for the optical selection rule.
		}
	\end{figure}

	To investigate the condensation behaviors, we performed self-consistent calculations based on the $k \cdot p$ model.
	Different from previous work~\cite{PhysRevLett.112.146405, PhysRevLett.120.186802, PhysRevLett.112.176403, doi:10.1126/sciadv.aau6120}, we only added the interaction in the electron-hole channel after diagonalizing Eq.~\ref{kp}, as the others are already included in the $GW$ calculations.
	In this way, the mean-field Hamiltonian reads:
	\begin{equation}
		\hat{H}_{\mathrm{MF}}=\sum_{\boldsymbol{k}} \hat{\Psi}^{\dagger}(\boldsymbol{k}) 	\mathcal{H}_{\text{MF}}(\boldsymbol{k}) \hat{\Psi}(\boldsymbol{k}),
	\end{equation}
	where
	\begin{equation}\label{mf}
		\mathcal{H}_{\text{MF}}(\boldsymbol{k})=\left(\begin{array}{cccc}
			\epsilon_{c, \uparrow}(\boldsymbol{k}) & 0 & \Delta_{\uparrow \uparrow}(\boldsymbol{k}) & \Delta_{\uparrow \downarrow}(\boldsymbol{k}) \\
			0 & \epsilon_{c, \downarrow}(\boldsymbol{k}) & \Delta_{\downarrow \uparrow}(\boldsymbol{k}) & \Delta_{\downarrow \downarrow}(\boldsymbol{k}) \\
			\Delta^*_{\uparrow \uparrow}(\boldsymbol{k}) & \Delta^*_{\downarrow \uparrow}(\boldsymbol{k}) & \epsilon_{v, \uparrow}(\boldsymbol{k}) & 0 \\
			\Delta^*_{\uparrow \downarrow}(\boldsymbol{k}) & \Delta^*_{\downarrow \downarrow}(\boldsymbol{k}) & 0 & \epsilon_{v, \downarrow}(\boldsymbol{k})
		\end{array}\right).
	\end{equation}
	The $\Delta(\boldsymbol{k})$s represent the excitonic order parameters, defined as:
	\begin{equation}\label{delta}
		\Delta_{\alpha \beta}(\boldsymbol{k})=-\sum_{\boldsymbol{k}^{\prime}} U_{\alpha \beta}(\boldsymbol{k}, \boldsymbol{k}^{\prime})\langle\hat{\Psi}^{\dagger}_{\beta}(\boldsymbol{k}^{\prime}) \hat{\Psi}_{\alpha}(\boldsymbol{k}^{\prime}) \rangle ,
	\end{equation}
	$\alpha \in[1,2]$ and $\beta \in[3,4] ;$ or $\alpha \in[3,4]$ and $\beta \in[1,2]$.
	$U_{\alpha \beta}(\boldsymbol{k}, \boldsymbol{k}^{\prime})=\mathcal{K}^{\text{d}}_{\alpha \beta \boldsymbol{k}, \alpha \beta \boldsymbol{k}^{\prime}}$ is the screened Coulomb interaction $W(\boldsymbol{k}-\boldsymbol{k}^{\prime})$ dressed by the form factors, and \begin{equation}\label{dm}
		\langle\hat{\Psi}^{\dagger}_{\beta}(\boldsymbol{k}) \hat{\Psi}_{\alpha}(\boldsymbol{k}) \rangle=\sum_i^4 [\mathcal{U}]_{\alpha, i} f_{\text{F}}(E_{\boldsymbol{k}, i}) \left[\mathcal{U}^{\dagger}\right]_{i, \beta}.
	\end{equation}
	Here $\mathcal{U}$ is the unitary matrix that diagonalizes the $\mathcal{H}_{\text{MF}}$, $E_{\boldsymbol{k}, i}$s are the corresponding eigenvalues, and $f_{\text{F}}$
	is the Fermi-Dirac distribution function.
	Eqs.~6-9 form a self-consistent loop to be solved.

	\begin{figure}[h]
		\includegraphics[width=1\linewidth]{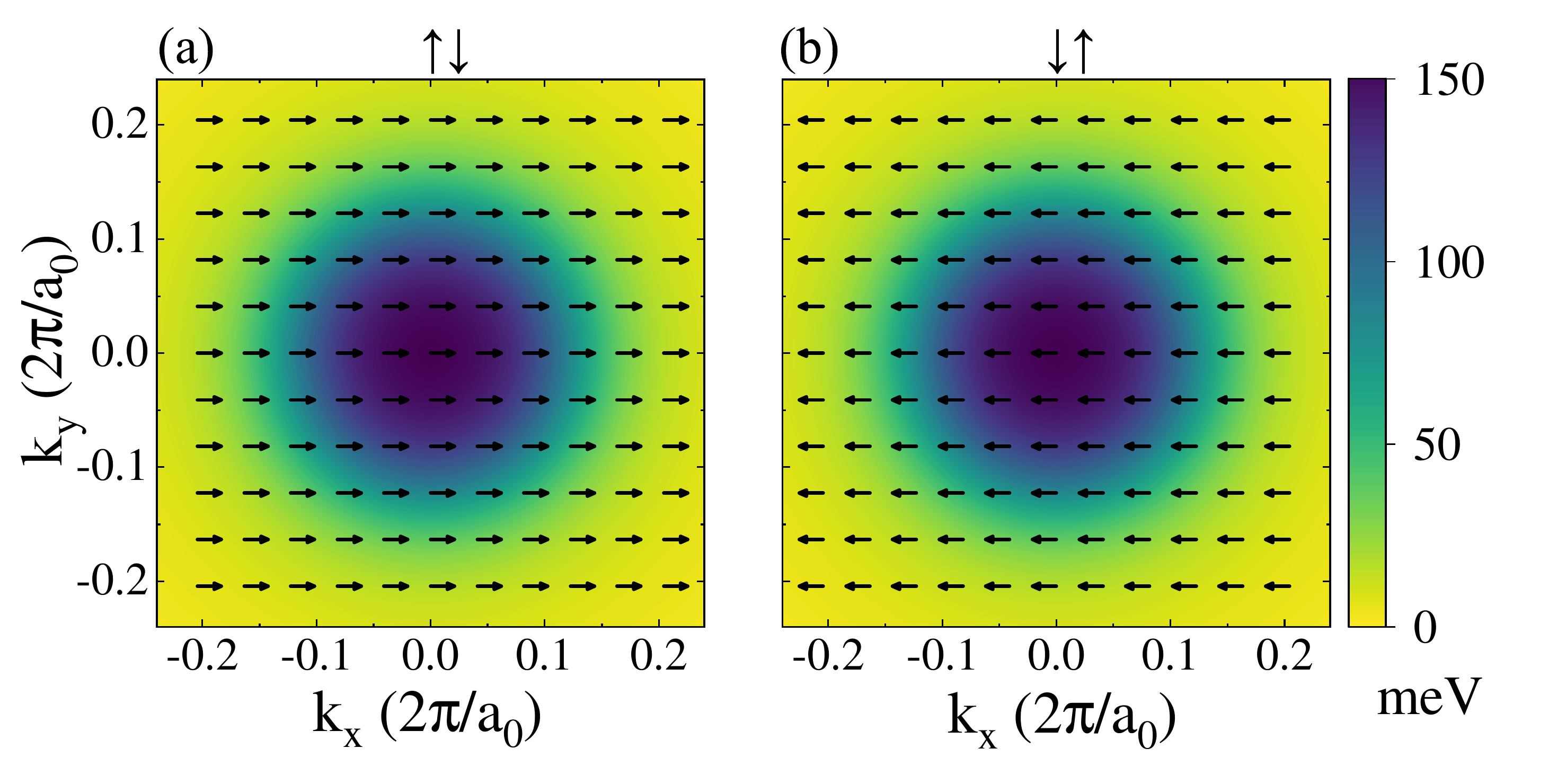}
		\caption{ \label{fig3}
			The absolute value and phase winding pattern of the order parameters in  (a) $\uparrow \downarrow$ and (b) $\downarrow \uparrow$ channels  at 0~K in the reciprocal space.
			The direction of the arrow denotes the phase angle and the colormap indicates the magnitude.
			The absolute value of the order parameters in $\uparrow \uparrow$ and $\downarrow \downarrow$ channels are zero and not shown here.}
	\end{figure}

	In Fig.~3, we show the self-consistent results at 0~K.
	Only the unlike-spin order parameters ($\Delta_{\uparrow \downarrow}(\boldsymbol{k}), \Delta_{\downarrow \uparrow}(\boldsymbol{k})$) are nonzero.
	This spin-triplet feature of the order parameters is consistent with the fact that $U_{\uparrow \downarrow}$ and  $U_{\downarrow \uparrow }$ are slightly larger than $U_{\uparrow \uparrow}$ and $U_{\downarrow \downarrow }$ in Eq.~\ref{delta}, and that the binding energy of the unlike-spin excitons are larger than the like-spin ones.
	Among all the self-consistent solutions,  s-wave like pairing is the most energetically favorable one.
	Therefore the absolute values of $\Delta_{\uparrow \downarrow}(\boldsymbol{k})$ and $\Delta_{\downarrow \uparrow}(\boldsymbol{k})$ are isotropic in $k$-space, with maximum at  $\boldsymbol{k}=0$, showing s-wave characteristics.
	The localization of the order parameters at the origin also means that the $k \cdot p$ model is rational.
	The phases of $\Delta_{\uparrow \downarrow}(\boldsymbol{k})$ and $\Delta_{\downarrow \uparrow}(\boldsymbol{k})$ remain constant all over the $k$-space.
	Due to the large order parameter in spin-triplet channel, the bandstructures in this EI phase are distinctly different from the original ones (Fig.~4(a)).
	A prominent increase of the band gap emerges, the conduction band flattens, and the valence band distort strongly, making the gap indirect.

	\begin{figure}[h]
		\includegraphics[width=1.2\linewidth]{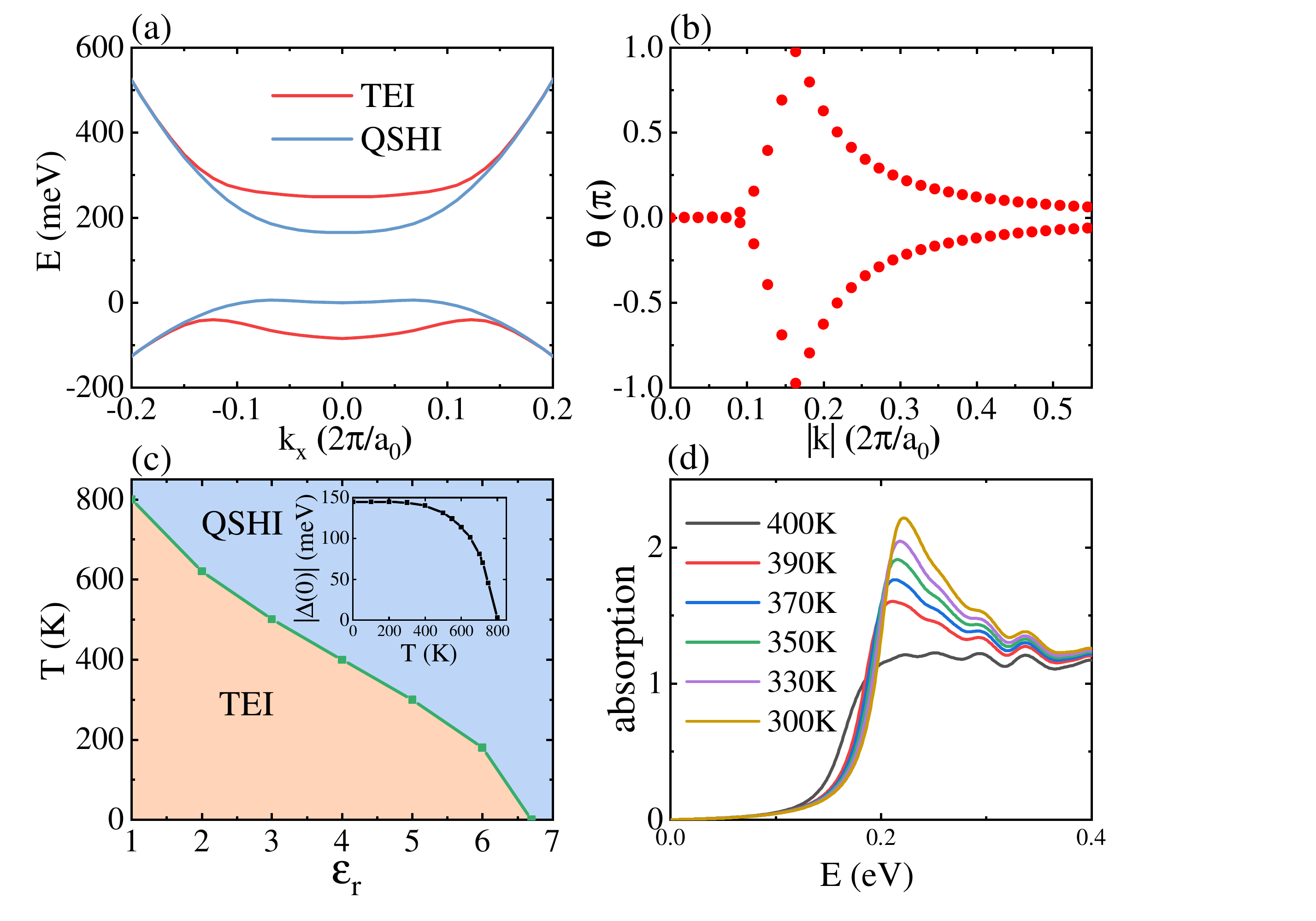}
		\caption{ \label{fig4}
			(a) Bandstructures of the excitonic phase (red lines) and original phase (blue lines).
			(b) Results of Wilson-loop calculation for the excitonic phase, where the red dots represent the phase angles of the two occupied bands.
			(c) The excitonic phase diagram with respect to different substrate dielectric constants and $T$s.
			The inset shows the dependence on the $T$ for the absolute value of order parmeter when the substrate dielectric constant equals unity.
			(d) Interband absorption spectrum at different $T$s when $\epsilon_r=4$.
		}
	\end{figure}

	To identify the symmetry in the excitonic phase, we rotate the converged $\mathcal{H}_{\text{MF}}$ back to the original basis in $\mathcal{H}_{0}$.
	The EI phase preserves the inversion symmetry $I$ and time-reversal symmetry $\mathcal{T}$, and the bands are still two-fold degenerate everywhere in the $k$-space due to the Kramers' theorem.
	During the formation of the EI phase, $\mathcal{T}$ preserves and the gap remains finite.
	Therefore the topology should be identical with the non-interacting case.
	To verify this point, we performed the Wilson loop calculation for the two degenerate occupied bands~\cite{PhysRevB.84.075119, doi:10.1080/00018732.2015.1068524}.
	The 2×2 overlap matrix $F_{i, i+1}^{m, n}=\left\langle u_i^m \mid u_{i+1}^n\right\rangle$ along square loops around $\Gamma$, where $m, n \in \{1, 2\}$ and $i$ is the index along the loop, are calculated~\cite{PhysRevLett.120.186802}.
	The phase angles $\theta_k$ corresponding to the eigenvalues of the product of all $F$s are shown in Fig.~4(b).
	Each occupied band winds the cylinder once in opposite directions, indicating that the nontrivial Z$_2$ topology survives in the excitonic phase.
	Majorana edge modes donot exist in this TEI phase, as the excitonic pairings are distinct from superconducting pairings~\cite{WOS:000455595400033}.
	Nevertheless, the topological excitonic edge states in the TEI phase may have novel properties in achieving and manipulating the Majorana modes through proximity effects, which deserve further investigation~\cite{WOS:000305907700002, PhysRevLett.100.096407, PhysRevB.86.121101}.
	On realistic substrates, screening can be estimated by substituting $W(\boldsymbol{q})$ with $W^{\prime}(\boldsymbol{q})=(2 \pi)/\left[|q|\left(\epsilon+2 \pi |q| \alpha_{2 \mathrm{D}}\right)\right]$ where $\epsilon = (1+\epsilon_{\text{r}})/2$ ($\epsilon_{\text{r}}$ is the dielectric constant of the substrate).
	As depicted in Fig.~4(c), $T_\text{C}$ of this exotic TEI phase reaches $\sim$800~K in the freestanding case ($\epsilon_{\text{r}}=1$).
	When $\epsilon_{\text{r}}$ increases, $T_\text{C}$ decreases gradually and finally drops to 0~K at $\epsilon_{\text{r}}=6.7$.
	These results evidently show that this TEI phase exists in a large phase space.
	In this TEI phase, there's no charge density wave as the excitons condensate at $Q=0$.
	Ferroelecticity does not exist either due to the same parity of the CBM and VBM.
	However, experimentally, ARPES and transport measurements are able to detect the bandstructures changes and topological excitonic edge states.
	Besides these, we also note that there are exotic optical response corresponding to the excitonic collective modes as predicted in
	recent studies~\cite{murakami2020collective, golevz2020nonlinear, tanabe2021third,kaneko2021bulk}.
	To demonstrate this, we calculate the optical absorption, i.e. the real part of the optical conductivity, through the Kubo formula:
	\begin{equation}
		\sigma(\omega)=\sum_{m \in c, n \in v} \int \frac{d \boldsymbol{q}}{(2 \pi)^2} \frac{4 i e^2 \omega \mathcal{O}_{\boldsymbol{q}}^{n, m}\left(f_{\text{F}}(E_{\boldsymbol{q}, n})-f_{\text{F}}(E_{\boldsymbol{q}, m})\right)}{\hbar \omega+i 0^{+}+E_{\boldsymbol{q}, n}-E_{\boldsymbol{q}, m}},
	\end{equation}
	\begin{equation}
		\mathcal{O}_{\mathbf{q}}^{n, m}=\frac{\hbar^2\left|\left\langle n \boldsymbol{q}\left|\hat{v}^x\right| m \boldsymbol{q}\right\rangle\right|^2}{\left(E_{\boldsymbol{q}, n}-E_{\boldsymbol{q}, m}\right)^2},
	\end{equation}
	and show the results at $\epsilon_{r}=4$ (like on h-BN substrate) in Fig.~4(d).
	When T drops below the $T_\text{C}=400$~K, the absorption spectrum shows a abrupt change that a peak appears at the band gap.
	This is due to the renormalization of the single-particle Hamiltonian Eq.~\ref{kp} from the EI order parameters $\Delta$.
	The peak grows more and more sharp as $T$ decreases further, and the absorption edge blueshifts as the $|\Delta|$ increases in the excitonic phase.
	Therefore, this TEI phase is not only achievable but also observable in realistic experimental setups.
	These unusual properties may have applications in different electronic, spintronic, and photonic devices.
	We hope this combined use of QSHIs with EI may stimulate more theoretical and experimental studies in such interdisciplinary areas.
	Note added: After we finished the present letter, we became aware of a related paper, Ref.~\onlinecite{dong2023robust}, whose results are consistent with ours in the finite overlapping part.

	\begin{acknowledgments}
		The DFT calculations and many-body $GW$/BSE calculations are performed using QUANTUM ESPRESSO package~\cite{giannozzi2009quantum} and YAMBO code~\cite{sangalli2019many} respectively.
		We are supported by the National Science Foundation of China under Grant Nos. 12234001 and 11934003, the Beijing Natural Science Foundation under Grant No. Z200004,
		the National Basic Research Programs of China under Grand Nos. 2022YFA1403503 and 2021YFA1400503,  the Strategic Priority Research Program of the Chinese Academy of Sciences Grant No. XDB33010400.
		The computational resources were provided by the supercomputer center in Peking University, China.
	\end{acknowledgments}
	

\begin{thebibliography}{49}%
		\makeatletter
		\providecommand \@ifxundefined [1]{%
			\@ifx{#1\undefined}
		}%
		\providecommand \@ifnum [1]{%
			\ifnum #1\expandafter \@firstoftwo
			\else \expandafter \@secondoftwo
			\fi
		}%
		\providecommand \@ifx [1]{%
			\ifx #1\expandafter \@firstoftwo
			\else \expandafter \@secondoftwo
			\fi
		}%
		\providecommand \natexlab [1]{#1}%
		\providecommand \enquote  [1]{``#1''}%
		\providecommand \bibnamefont  [1]{#1}%
		\providecommand \bibfnamefont [1]{#1}%
		\providecommand \citenamefont [1]{#1}%
		\providecommand \href@noop [0]{\@secondoftwo}%
		\providecommand \href [0]{\begingroup \@sanitize@url \@href}%
		\providecommand \@href[1]{\@@startlink{#1}\@@href}%
		\providecommand \@@href[1]{\endgroup#1\@@endlink}%
		\providecommand \@sanitize@url [0]{\catcode `\\12\catcode `\$12\catcode
			`\&12\catcode `\#12\catcode `\^12\catcode `\_12\catcode `\%12\relax}%
		\providecommand \@@startlink[1]{}%
		\providecommand \@@endlink[0]{}%
		\providecommand \url  [0]{\begingroup\@sanitize@url \@url }%
		\providecommand \@url [1]{\endgroup\@href {#1}{\urlprefix }}%
		\providecommand \urlprefix  [0]{URL }%
		\providecommand \Eprint [0]{\href }%
		\providecommand \doibase [0]{http://dx.doi.org/}%
		\providecommand \selectlanguage [0]{\@gobble}%
		\providecommand \bibinfo  [0]{\@secondoftwo}%
		\providecommand \bibfield  [0]{\@secondoftwo}%
		\providecommand \translation [1]{[#1]}%
		\providecommand \BibitemOpen [0]{}%
		\providecommand \bibitemStop [0]{}%
		\providecommand \bibitemNoStop [0]{.\EOS\space}%
		\providecommand \EOS [0]{\spacefactor3000\relax}%
		\providecommand \BibitemShut  [1]{\csname bibitem#1\endcsname}%
		\let\auto@bib@innerbib\@empty
		\bibitem [{\citenamefont {Kane}\ and\ \citenamefont
			{Mele}(2005{\natexlab{a}})}]{PhysRevLett.95.226801}%
		\BibitemOpen
		\bibfield  {author} {\bibinfo {author} {\bibfnamefont {C.~L.}\ \bibnamefont
				{Kane}}\ and\ \bibinfo {author} {\bibfnamefont {E.~J.}\ \bibnamefont
				{Mele}},\ }\href {\doibase 10.1103/PhysRevLett.95.226801} {\bibfield
			{journal} {\bibinfo  {journal} {Phys. Rev. Lett.}\ }\textbf {\bibinfo
				{volume} {95}},\ \bibinfo {pages} {226801} (\bibinfo {year}
			{2005}{\natexlab{a}})}\BibitemShut {NoStop}%
		\bibitem [{\citenamefont {Kane}\ and\ \citenamefont
			{Mele}(2005{\natexlab{b}})}]{PhysRevLett.95.146802}%
		\BibitemOpen
		\bibfield  {author} {\bibinfo {author} {\bibfnamefont {C.~L.}\ \bibnamefont
				{Kane}}\ and\ \bibinfo {author} {\bibfnamefont {E.~J.}\ \bibnamefont
				{Mele}},\ }\href {\doibase 10.1103/PhysRevLett.95.146802} {\bibfield
			{journal} {\bibinfo  {journal} {Phys. Rev. Lett.}\ }\textbf {\bibinfo
				{volume} {95}},\ \bibinfo {pages} {146802} (\bibinfo {year}
			{2005}{\natexlab{b}})}\BibitemShut {NoStop}%
		\bibitem [{\citenamefont {Liu}\ \emph {et~al.}(2008)\citenamefont {Liu},
			\citenamefont {Hughes}, \citenamefont {Qi}, \citenamefont {Wang},\ and\
			\citenamefont {Zhang}}]{PhysRevLett.100.236601}%
		\BibitemOpen
		\bibfield  {author} {\bibinfo {author} {\bibfnamefont {C.}~\bibnamefont
				{Liu}}, \bibinfo {author} {\bibfnamefont {T.~L.}\ \bibnamefont {Hughes}},
			\bibinfo {author} {\bibfnamefont {X.-L.}\ \bibnamefont {Qi}}, \bibinfo
			{author} {\bibfnamefont {K.}~\bibnamefont {Wang}}, \ and\ \bibinfo {author}
			{\bibfnamefont {S.-C.}\ \bibnamefont {Zhang}},\ }\href {\doibase
			10.1103/PhysRevLett.100.236601} {\bibfield  {journal} {\bibinfo  {journal}
				{Phys. Rev. Lett.}\ }\textbf {\bibinfo {volume} {100}},\ \bibinfo {pages}
			{236601} (\bibinfo {year} {2008})}\BibitemShut {NoStop}%
		\bibitem [{\citenamefont {Bernevig}\ \emph {et~al.}(2006)\citenamefont
			{Bernevig}, \citenamefont {Hughes},\ and\ \citenamefont
			{Zhang}}]{doi:10.1126/science.1133734}%
		\BibitemOpen
		\bibfield  {author} {\bibinfo {author} {\bibfnamefont {B.~A.}\ \bibnamefont
				{Bernevig}}, \bibinfo {author} {\bibfnamefont {T.~L.}\ \bibnamefont
				{Hughes}}, \ and\ \bibinfo {author} {\bibfnamefont {S.-C.}\ \bibnamefont
				{Zhang}},\ }\href {\doibase 10.1126/science.1133734} {\bibfield  {journal}
			{\bibinfo  {journal} {Science}\ }\textbf {\bibinfo {volume} {314}},\ \bibinfo
			{pages} {1757} (\bibinfo {year} {2006})}\BibitemShut {NoStop}%
		\bibitem [{\citenamefont {Kou}\ \emph {et~al.}(2017)\citenamefont {Kou},
			\citenamefont {Ma}, \citenamefont {Sun}, \citenamefont {Heine},\ and\
			\citenamefont {Chen}}]{doi:10.1021/acs.jpclett.7b00222}%
		\BibitemOpen
		\bibfield  {author} {\bibinfo {author} {\bibfnamefont {L.}~\bibnamefont
				{Kou}}, \bibinfo {author} {\bibfnamefont {Y.}~\bibnamefont {Ma}}, \bibinfo
			{author} {\bibfnamefont {Z.}~\bibnamefont {Sun}}, \bibinfo {author}
			{\bibfnamefont {T.}~\bibnamefont {Heine}}, \ and\ \bibinfo {author}
			{\bibfnamefont {C.}~\bibnamefont {Chen}},\ }\href {\doibase
			10.1021/acs.jpclett.7b00222} {\bibfield  {journal} {\bibinfo  {journal} {J.
					Phys. Chem. Lett.}\ }\textbf {\bibinfo {volume} {8}},\ \bibinfo {pages}
			{1905} (\bibinfo {year} {2017})}\BibitemShut {NoStop}%
		\bibitem [{\citenamefont {Lodge}\ \emph {et~al.}(2021)\citenamefont {Lodge},
			\citenamefont {Yang}, \citenamefont {Mukherjee},\ and\ \citenamefont
			{Weber}}]{https://doi.org/10.1002/adma.202008029}%
		\BibitemOpen
		\bibfield  {author} {\bibinfo {author} {\bibfnamefont {M.~S.}\ \bibnamefont
				{Lodge}}, \bibinfo {author} {\bibfnamefont {S.~A.}\ \bibnamefont {Yang}},
			\bibinfo {author} {\bibfnamefont {S.}~\bibnamefont {Mukherjee}}, \ and\
			\bibinfo {author} {\bibfnamefont {B.}~\bibnamefont {Weber}},\ }\href
		{\doibase https://doi.org/10.1002/adma.202008029} {\bibfield  {journal}
			{\bibinfo  {journal} {Adv. Mater.}\ }\textbf {\bibinfo {volume} {33}},\
			\bibinfo {pages} {2008029} (\bibinfo {year} {2021})}\BibitemShut {NoStop}%
		\bibitem [{\citenamefont {Du}\ \emph {et~al.}(2017)\citenamefont {Du},
			\citenamefont {Li}, \citenamefont {Lou}, \citenamefont {Sullivan},
			\citenamefont {Chang}, \citenamefont {Kono},\ and\ \citenamefont
			{Du}}]{du2017evidence}%
		\BibitemOpen
		\bibfield  {author} {\bibinfo {author} {\bibfnamefont {L.}~\bibnamefont
				{Du}}, \bibinfo {author} {\bibfnamefont {X.}~\bibnamefont {Li}}, \bibinfo
			{author} {\bibfnamefont {W.}~\bibnamefont {Lou}}, \bibinfo {author}
			{\bibfnamefont {G.}~\bibnamefont {Sullivan}}, \bibinfo {author}
			{\bibfnamefont {K.}~\bibnamefont {Chang}}, \bibinfo {author} {\bibfnamefont
				{J.}~\bibnamefont {Kono}}, \ and\ \bibinfo {author} {\bibfnamefont {R.-R.}\
				\bibnamefont {Du}},\ }\href@noop {} {\bibfield  {journal} {\bibinfo
				{journal} {Nat. Commun.}\ }\textbf {\bibinfo {volume} {8}},\ \bibinfo {pages}
			{1} (\bibinfo {year} {2017})}\BibitemShut {NoStop}%
		\bibitem [{\citenamefont {Wang}\ \emph
			{et~al.}(2019{\natexlab{a}})\citenamefont {Wang}, \citenamefont {Rhodes},
			\citenamefont {Watanabe}, \citenamefont {Taniguchi}, \citenamefont {Hone},
			\citenamefont {Shan},\ and\ \citenamefont {Mak}}]{wang2019evidence}%
		\BibitemOpen
		\bibfield  {author} {\bibinfo {author} {\bibfnamefont {Z.}~\bibnamefont
				{Wang}}, \bibinfo {author} {\bibfnamefont {D.~A.}\ \bibnamefont {Rhodes}},
			\bibinfo {author} {\bibfnamefont {K.}~\bibnamefont {Watanabe}}, \bibinfo
			{author} {\bibfnamefont {T.}~\bibnamefont {Taniguchi}}, \bibinfo {author}
			{\bibfnamefont {J.~C.}\ \bibnamefont {Hone}}, \bibinfo {author}
			{\bibfnamefont {J.}~\bibnamefont {Shan}}, \ and\ \bibinfo {author}
			{\bibfnamefont {K.~F.}\ \bibnamefont {Mak}},\ }\href@noop {} {\bibfield
			{journal} {\bibinfo  {journal} {Nature}\ }\textbf {\bibinfo {volume} {574}},\
			\bibinfo {pages} {76} (\bibinfo {year} {2019}{\natexlab{a}})}\BibitemShut
		{NoStop}%
		\bibitem [{\citenamefont {Wakisaka}\ \emph {et~al.}(2009)\citenamefont
			{Wakisaka}, \citenamefont {Sudayama}, \citenamefont {Takubo}, \citenamefont
			{Mizokawa}, \citenamefont {Arita}, \citenamefont {Namatame}, \citenamefont
			{Taniguchi}, \citenamefont {Katayama}, \citenamefont {Nohara},\ and\
			\citenamefont {Takagi}}]{wakisaka2009excitonic}%
		\BibitemOpen
		\bibfield  {author} {\bibinfo {author} {\bibfnamefont {Y.}~\bibnamefont
				{Wakisaka}}, \bibinfo {author} {\bibfnamefont {T.}~\bibnamefont {Sudayama}},
			\bibinfo {author} {\bibfnamefont {K.}~\bibnamefont {Takubo}}, \bibinfo
			{author} {\bibfnamefont {T.}~\bibnamefont {Mizokawa}}, \bibinfo {author}
			{\bibfnamefont {M.}~\bibnamefont {Arita}}, \bibinfo {author} {\bibfnamefont
				{H.}~\bibnamefont {Namatame}}, \bibinfo {author} {\bibfnamefont
				{M.}~\bibnamefont {Taniguchi}}, \bibinfo {author} {\bibfnamefont
				{N.}~\bibnamefont {Katayama}}, \bibinfo {author} {\bibfnamefont
				{M.}~\bibnamefont {Nohara}}, \ and\ \bibinfo {author} {\bibfnamefont
				{H.}~\bibnamefont {Takagi}},\ }\href@noop {} {\bibfield  {journal} {\bibinfo
				{journal} {Phys. Rev. Lett.}\ }\textbf {\bibinfo {volume} {103}},\ \bibinfo
			{pages} {026402} (\bibinfo {year} {2009})}\BibitemShut {NoStop}%
		\bibitem [{\citenamefont {Lu}\ \emph {et~al.}(2017)\citenamefont {Lu},
			\citenamefont {Kono}, \citenamefont {Larkin}, \citenamefont {Rost},
			\citenamefont {Takayama}, \citenamefont {Boris}, \citenamefont {Keimer},\
			and\ \citenamefont {Takagi}}]{lu2017zero}%
		\BibitemOpen
		\bibfield  {author} {\bibinfo {author} {\bibfnamefont {Y.~F.}\ \bibnamefont
				{Lu}}, \bibinfo {author} {\bibfnamefont {H.}~\bibnamefont {Kono}}, \bibinfo
			{author} {\bibfnamefont {T.~I.}\ \bibnamefont {Larkin}}, \bibinfo {author}
			{\bibfnamefont {A.~W.}\ \bibnamefont {Rost}}, \bibinfo {author}
			{\bibfnamefont {T.}~\bibnamefont {Takayama}}, \bibinfo {author}
			{\bibfnamefont {A.~V.}\ \bibnamefont {Boris}}, \bibinfo {author}
			{\bibfnamefont {B.}~\bibnamefont {Keimer}}, \ and\ \bibinfo {author}
			{\bibfnamefont {H.}~\bibnamefont {Takagi}},\ }\href@noop {} {\bibfield
			{journal} {\bibinfo  {journal} {Nat. Commun.}\ }\textbf {\bibinfo {volume}
				{8}},\ \bibinfo {pages} {1} (\bibinfo {year} {2017})}\BibitemShut {NoStop}%
		\bibitem [{\citenamefont {Mor}\ \emph {et~al.}(2017)\citenamefont {Mor},
			\citenamefont {Herzog}, \citenamefont {Gole\ifmmode~\check{z}\else
				\v{z}\fi{}}, \citenamefont {Werner}, \citenamefont {Eckstein}, \citenamefont
			{Katayama}, \citenamefont {Nohara}, \citenamefont {Takagi}, \citenamefont
			{Mizokawa}, \citenamefont {Monney},\ and\ \citenamefont
			{St\"ahler}}]{mor2017ultrafast}%
		\BibitemOpen
		\bibfield  {author} {\bibinfo {author} {\bibfnamefont {S.}~\bibnamefont
				{Mor}}, \bibinfo {author} {\bibfnamefont {M.}~\bibnamefont {Herzog}},
			\bibinfo {author} {\bibfnamefont {D.}~\bibnamefont
				{Gole\ifmmode~\check{z}\else \v{z}\fi{}}}, \bibinfo {author} {\bibfnamefont
				{P.}~\bibnamefont {Werner}}, \bibinfo {author} {\bibfnamefont
				{M.}~\bibnamefont {Eckstein}}, \bibinfo {author} {\bibfnamefont
				{N.}~\bibnamefont {Katayama}}, \bibinfo {author} {\bibfnamefont
				{M.}~\bibnamefont {Nohara}}, \bibinfo {author} {\bibfnamefont
				{H.}~\bibnamefont {Takagi}}, \bibinfo {author} {\bibfnamefont
				{T.}~\bibnamefont {Mizokawa}}, \bibinfo {author} {\bibfnamefont
				{C.}~\bibnamefont {Monney}}, \ and\ \bibinfo {author} {\bibfnamefont
				{J.}~\bibnamefont {St\"ahler}},\ }\href {\doibase
			10.1103/PhysRevLett.119.086401} {\bibfield  {journal} {\bibinfo  {journal}
				{Phys. Rev. Lett.}\ }\textbf {\bibinfo {volume} {119}},\ \bibinfo {pages}
			{086401} (\bibinfo {year} {2017})}\BibitemShut {NoStop}%
		\bibitem [{\citenamefont {Jiang}\ \emph {et~al.}(2018)\citenamefont {Jiang},
			\citenamefont {Li}, \citenamefont {Zhang},\ and\ \citenamefont
			{Duan}}]{jiang2018realizing}%
		\BibitemOpen
		\bibfield  {author} {\bibinfo {author} {\bibfnamefont {Z.}~\bibnamefont
				{Jiang}}, \bibinfo {author} {\bibfnamefont {Y.}~\bibnamefont {Li}}, \bibinfo
			{author} {\bibfnamefont {S.}~\bibnamefont {Zhang}}, \ and\ \bibinfo {author}
			{\bibfnamefont {W.}~\bibnamefont {Duan}},\ }\href {\doibase
			10.1103/PhysRevB.98.081408} {\bibfield  {journal} {\bibinfo  {journal} {Phys.
					Rev. B}\ }\textbf {\bibinfo {volume} {98}},\ \bibinfo {pages} {081408}
			(\bibinfo {year} {2018})}\BibitemShut {NoStop}%
		\bibitem [{\citenamefont {Jiang}\ \emph {et~al.}(2019)\citenamefont {Jiang},
			\citenamefont {Li}, \citenamefont {Duan},\ and\ \citenamefont
			{Zhang}}]{jiang2019half}%
		\BibitemOpen
		\bibfield  {author} {\bibinfo {author} {\bibfnamefont {Z.}~\bibnamefont
				{Jiang}}, \bibinfo {author} {\bibfnamefont {Y.}~\bibnamefont {Li}}, \bibinfo
			{author} {\bibfnamefont {W.}~\bibnamefont {Duan}}, \ and\ \bibinfo {author}
			{\bibfnamefont {S.}~\bibnamefont {Zhang}},\ }\href@noop {} {\bibfield
			{journal} {\bibinfo  {journal} {Phys. Rev. Lett.}\ }\textbf {\bibinfo
				{volume} {122}},\ \bibinfo {pages} {236402} (\bibinfo {year}
			{2019})}\BibitemShut {NoStop}%
		\bibitem [{\citenamefont {Jiang}\ \emph {et~al.}(2020)\citenamefont {Jiang},
			\citenamefont {Lou}, \citenamefont {Liu}, \citenamefont {Li}, \citenamefont
			{Song}, \citenamefont {Chang}, \citenamefont {Duan},\ and\ \citenamefont
			{Zhang}}]{jiang2020spin}%
		\BibitemOpen
		\bibfield  {author} {\bibinfo {author} {\bibfnamefont {Z.}~\bibnamefont
				{Jiang}}, \bibinfo {author} {\bibfnamefont {W.}~\bibnamefont {Lou}}, \bibinfo
			{author} {\bibfnamefont {Y.}~\bibnamefont {Liu}}, \bibinfo {author}
			{\bibfnamefont {Y.}~\bibnamefont {Li}}, \bibinfo {author} {\bibfnamefont
				{H.}~\bibnamefont {Song}}, \bibinfo {author} {\bibfnamefont {K.}~\bibnamefont
				{Chang}}, \bibinfo {author} {\bibfnamefont {W.}~\bibnamefont {Duan}}, \ and\
			\bibinfo {author} {\bibfnamefont {S.}~\bibnamefont {Zhang}},\ }\href@noop {}
		{\bibfield  {journal} {\bibinfo  {journal} {Phys. Rev. Lett.}\ }\textbf
			{\bibinfo {volume} {124}},\ \bibinfo {pages} {166401} (\bibinfo {year}
			{2020})}\BibitemShut {NoStop}%
		\bibitem [{\citenamefont {Varsano}\ \emph {et~al.}(2020)\citenamefont
			{Varsano}, \citenamefont {Palummo}, \citenamefont {Molinari},\ and\
			\citenamefont {Rontani}}]{Varsano2020367}%
		\BibitemOpen
		\bibfield  {author} {\bibinfo {author} {\bibfnamefont {D.}~\bibnamefont
				{Varsano}}, \bibinfo {author} {\bibfnamefont {M.}~\bibnamefont {Palummo}},
			\bibinfo {author} {\bibfnamefont {E.}~\bibnamefont {Molinari}}, \ and\
			\bibinfo {author} {\bibfnamefont {M.}~\bibnamefont {Rontani}},\ }\href
		{\doibase 10.1038/s41565-020-0650-4} {\bibfield  {journal} {\bibinfo
				{journal} {Nat. Nanotechnol.}\ }\textbf {\bibinfo {volume} {15}},\ \bibinfo
			{pages} {367} (\bibinfo {year} {2020})}\BibitemShut {NoStop}%
		\bibitem [{\citenamefont {Yang}\ \emph {et~al.}(2022)\citenamefont {Yang},
			\citenamefont {Wang},\ and\ \citenamefont {Li}}]{Yang_2022}%
		\BibitemOpen
		\bibfield  {author} {\bibinfo {author} {\bibfnamefont {H.}~\bibnamefont
				{Yang}}, \bibinfo {author} {\bibfnamefont {X.}~\bibnamefont {Wang}}, \ and\
			\bibinfo {author} {\bibfnamefont {X.-Z.}\ \bibnamefont {Li}},\ }\href
		{\doibase 10.1088/1367-2630/ac81e4} {\bibfield  {journal} {\bibinfo
				{journal} {New J. Phys.}\ }\textbf {\bibinfo {volume} {24}},\ \bibinfo
			{pages} {083010} (\bibinfo {year} {2022})}\BibitemShut {NoStop}%
		\bibitem [{\citenamefont {Keldysh}\ and\ \citenamefont
			{Kopaev}(1965)}]{keldysh1965possible}%
		\BibitemOpen
		\bibfield  {author} {\bibinfo {author} {\bibfnamefont {L.~V.}\ \bibnamefont
				{Keldysh}}\ and\ \bibinfo {author} {\bibfnamefont {Y.~V.}\ \bibnamefont
				{Kopaev}},\ }\href@noop {} {\bibfield  {journal} {\bibinfo  {journal} {Sov.
					Phys. Solid State}\ }\textbf {\bibinfo {volume} {6}},\ \bibinfo {pages}
			{2219} (\bibinfo {year} {1965})}\BibitemShut {NoStop}%
		\bibitem [{\citenamefont {Kohn}(1967)}]{kohn1967excitonic}%
		\BibitemOpen
		\bibfield  {author} {\bibinfo {author} {\bibfnamefont {W.}~\bibnamefont
				{Kohn}},\ }\href@noop {} {\bibfield  {journal} {\bibinfo  {journal} {Phys.
					Rev. Lett.}\ }\textbf {\bibinfo {volume} {19}},\ \bibinfo {pages} {439}
			(\bibinfo {year} {1967})}\BibitemShut {NoStop}%
		\bibitem [{\citenamefont {J{\'e}rome}\ \emph {et~al.}(1967)\citenamefont
			{J{\'e}rome}, \citenamefont {Rice},\ and\ \citenamefont
			{Kohn}}]{jerome1967excitonic}%
		\BibitemOpen
		\bibfield  {author} {\bibinfo {author} {\bibfnamefont {D.}~\bibnamefont
				{J{\'e}rome}}, \bibinfo {author} {\bibfnamefont {T.~M.}\ \bibnamefont
				{Rice}}, \ and\ \bibinfo {author} {\bibfnamefont {W.}~\bibnamefont {Kohn}},\
		}\href@noop {} {\bibfield  {journal} {\bibinfo  {journal} {Phys. Rev.}\
			}\textbf {\bibinfo {volume} {158}},\ \bibinfo {pages} {462} (\bibinfo {year}
			{1967})}\BibitemShut {NoStop}%
		\bibitem [{\citenamefont {Halperin}\ and\ \citenamefont
			{Rice}(1968)}]{RevModPhys.40.755}%
		\BibitemOpen
		\bibfield  {author} {\bibinfo {author} {\bibfnamefont {B.~I.}\ \bibnamefont
				{Halperin}}\ and\ \bibinfo {author} {\bibfnamefont {T.~M.}\ \bibnamefont
				{Rice}},\ }\href {\doibase 10.1103/RevModPhys.40.755} {\bibfield  {journal}
			{\bibinfo  {journal} {Rev. Mod. Phys.}\ }\textbf {\bibinfo {volume} {40}},\
			\bibinfo {pages} {755} (\bibinfo {year} {1968})}\BibitemShut {NoStop}%
		\bibitem [{\citenamefont {Cercellier}\ \emph {et~al.}(2007)\citenamefont
			{Cercellier}, \citenamefont {Monney}, \citenamefont {Clerc}, \citenamefont
			{Battaglia}, \citenamefont {Despont}, \citenamefont {Garnier}, \citenamefont
			{Beck}, \citenamefont {Aebi}, \citenamefont {Patthey}, \citenamefont
			{Berger},\ and\ \citenamefont {Forr\'o}}]{cercellier2007evidence}%
		\BibitemOpen
		\bibfield  {author} {\bibinfo {author} {\bibfnamefont {H.}~\bibnamefont
				{Cercellier}}, \bibinfo {author} {\bibfnamefont {C.}~\bibnamefont {Monney}},
			\bibinfo {author} {\bibfnamefont {F.}~\bibnamefont {Clerc}}, \bibinfo
			{author} {\bibfnamefont {C.}~\bibnamefont {Battaglia}}, \bibinfo {author}
			{\bibfnamefont {L.}~\bibnamefont {Despont}}, \bibinfo {author} {\bibfnamefont
				{M.~G.}\ \bibnamefont {Garnier}}, \bibinfo {author} {\bibfnamefont
				{H.}~\bibnamefont {Beck}}, \bibinfo {author} {\bibfnamefont {P.}~\bibnamefont
				{Aebi}}, \bibinfo {author} {\bibfnamefont {L.}~\bibnamefont {Patthey}},
			\bibinfo {author} {\bibfnamefont {H.}~\bibnamefont {Berger}}, \ and\ \bibinfo
			{author} {\bibfnamefont {L.}~\bibnamefont {Forr\'o}},\ }\href {\doibase
			10.1103/PhysRevLett.99.146403} {\bibfield  {journal} {\bibinfo  {journal}
				{Phys. Rev. Lett.}\ }\textbf {\bibinfo {volume} {99}},\ \bibinfo {pages}
			{146403} (\bibinfo {year} {2007})}\BibitemShut {NoStop}%
		\bibitem [{\citenamefont {Kogar}\ \emph {et~al.}(2017)\citenamefont {Kogar}
			\emph {et~al.}}]{kogar2017signatures}%
		\BibitemOpen
		\bibfield  {author} {\bibinfo {author} {\bibfnamefont {A.}~\bibnamefont
				{Kogar}} \emph {et~al.},\ }\href@noop {} {\bibfield  {journal} {\bibinfo
				{journal} {Science}\ }\textbf {\bibinfo {volume} {358}},\ \bibinfo {pages}
			{1314} (\bibinfo {year} {2017})}\BibitemShut {NoStop}%
		\bibitem [{\citenamefont {Sun}\ \emph {et~al.}(2022)\citenamefont {Sun} \emph
			{et~al.}}]{sun2022evidence}%
		\BibitemOpen
		\bibfield  {author} {\bibinfo {author} {\bibfnamefont {B.}~\bibnamefont
				{Sun}} \emph {et~al.},\ }\href@noop {} {\bibfield  {journal} {\bibinfo
				{journal} {Nat. Phys.}\ }\textbf {\bibinfo {volume} {18}},\ \bibinfo {pages}
			{94} (\bibinfo {year} {2022})}\BibitemShut {NoStop}%
		\bibitem [{\citenamefont {Jia}\ \emph {et~al.}(2022)\citenamefont {Jia} \emph
			{et~al.}}]{jia2022evidence}%
		\BibitemOpen
		\bibfield  {author} {\bibinfo {author} {\bibfnamefont {Y.}~\bibnamefont
				{Jia}} \emph {et~al.},\ }\href@noop {} {\bibfield  {journal} {\bibinfo
				{journal} {Nat. Phys.}\ }\textbf {\bibinfo {volume} {18}},\ \bibinfo {pages}
			{87} (\bibinfo {year} {2022})}\BibitemShut {NoStop}%
		\bibitem [{\citenamefont {Jiang}\ \emph {et~al.}(2017)\citenamefont {Jiang},
			\citenamefont {Liu}, \citenamefont {Li},\ and\ \citenamefont
			{Duan}}]{jiang2017scaling}%
		\BibitemOpen
		\bibfield  {author} {\bibinfo {author} {\bibfnamefont {Z.}~\bibnamefont
				{Jiang}}, \bibinfo {author} {\bibfnamefont {Z.}~\bibnamefont {Liu}}, \bibinfo
			{author} {\bibfnamefont {Y.}~\bibnamefont {Li}}, \ and\ \bibinfo {author}
			{\bibfnamefont {W.}~\bibnamefont {Duan}},\ }\href@noop {} {\bibfield
			{journal} {\bibinfo  {journal} {Phys. Rev. Lett.}\ }\textbf {\bibinfo
				{volume} {118}},\ \bibinfo {pages} {266401} (\bibinfo {year}
			{2017})}\BibitemShut {NoStop}%
		\bibitem [{\citenamefont {Si}\ \emph {et~al.}(2016)\citenamefont {Si},
			\citenamefont {Jin}, \citenamefont {Zhou}, \citenamefont {Sun},\ and\
			\citenamefont {Liu}}]{doi:10.1021/acs.nanolett.6b03118}%
		\BibitemOpen
		\bibfield  {author} {\bibinfo {author} {\bibfnamefont {C.}~\bibnamefont
				{Si}}, \bibinfo {author} {\bibfnamefont {K.-H.}\ \bibnamefont {Jin}},
			\bibinfo {author} {\bibfnamefont {J.}~\bibnamefont {Zhou}}, \bibinfo {author}
			{\bibfnamefont {Z.}~\bibnamefont {Sun}}, \ and\ \bibinfo {author}
			{\bibfnamefont {F.}~\bibnamefont {Liu}},\ }\href {\doibase
			10.1021/acs.nanolett.6b03118} {\bibfield  {journal} {\bibinfo  {journal}
				{Nano Lett.}\ }\textbf {\bibinfo {volume} {16}},\ \bibinfo {pages} {6584}
			(\bibinfo {year} {2016})}\BibitemShut {NoStop}%
		\bibitem [{\citenamefont {Wang}\ \emph {et~al.}(2017)\citenamefont {Wang},
			\citenamefont {Ji}, \citenamefont {Zhang}, \citenamefont {Li}, \citenamefont
			{Zhang}, \citenamefont {Wang}, \citenamefont {Li},\ and\ \citenamefont
			{Yan}}]{doi:10.1063/1.4983781}%
		\BibitemOpen
		\bibfield  {author} {\bibinfo {author} {\bibfnamefont {Y.}~\bibnamefont
				{Wang}}, \bibinfo {author} {\bibfnamefont {W.}~\bibnamefont {Ji}}, \bibinfo
			{author} {\bibfnamefont {C.}~\bibnamefont {Zhang}}, \bibinfo {author}
			{\bibfnamefont {P.}~\bibnamefont {Li}}, \bibinfo {author} {\bibfnamefont
				{S.}~\bibnamefont {Zhang}}, \bibinfo {author} {\bibfnamefont
				{P.}~\bibnamefont {Wang}}, \bibinfo {author} {\bibfnamefont {S.}~\bibnamefont
				{Li}}, \ and\ \bibinfo {author} {\bibfnamefont {S.}~\bibnamefont {Yan}},\
		}\href {\doibase 10.1063/1.4983781} {\bibfield  {journal} {\bibinfo
				{journal} {Appl. Phys. Lett.}\ }\textbf {\bibinfo {volume} {110}},\ \bibinfo
			{pages} {213101} (\bibinfo {year} {2017})}\BibitemShut {NoStop}%
		\bibitem [{\citenamefont {Budich}\ \emph {et~al.}(2014)\citenamefont {Budich},
			\citenamefont {Trauzettel},\ and\ \citenamefont
			{Michetti}}]{PhysRevLett.112.146405}%
		\BibitemOpen
		\bibfield  {author} {\bibinfo {author} {\bibfnamefont {J.~C.}\ \bibnamefont
				{Budich}}, \bibinfo {author} {\bibfnamefont {B.}~\bibnamefont {Trauzettel}},
			\ and\ \bibinfo {author} {\bibfnamefont {P.}~\bibnamefont {Michetti}},\
		}\href {\doibase 10.1103/PhysRevLett.112.146405} {\bibfield  {journal}
			{\bibinfo  {journal} {Phys. Rev. Lett.}\ }\textbf {\bibinfo {volume} {112}},\
			\bibinfo {pages} {146405} (\bibinfo {year} {2014})}\BibitemShut {NoStop}%
		\bibitem [{\citenamefont {Pikulin}\ and\ \citenamefont
			{Hyart}(2014)}]{PhysRevLett.112.176403}%
		\BibitemOpen
		\bibfield  {author} {\bibinfo {author} {\bibfnamefont {D.~I.}\ \bibnamefont
				{Pikulin}}\ and\ \bibinfo {author} {\bibfnamefont {T.}~\bibnamefont
				{Hyart}},\ }\href {\doibase 10.1103/PhysRevLett.112.176403} {\bibfield
			{journal} {\bibinfo  {journal} {Phys. Rev. Lett.}\ }\textbf {\bibinfo
				{volume} {112}},\ \bibinfo {pages} {176403} (\bibinfo {year}
			{2014})}\BibitemShut {NoStop}%
		\bibitem [{\citenamefont {Xue}\ and\ \citenamefont
			{MacDonald}(2018)}]{PhysRevLett.120.186802}%
		\BibitemOpen
		\bibfield  {author} {\bibinfo {author} {\bibfnamefont {F.}~\bibnamefont
				{Xue}}\ and\ \bibinfo {author} {\bibfnamefont {A.~H.}\ \bibnamefont
				{MacDonald}},\ }\href {\doibase 10.1103/PhysRevLett.120.186802} {\bibfield
			{journal} {\bibinfo  {journal} {Phys. Rev. Lett.}\ }\textbf {\bibinfo
				{volume} {120}},\ \bibinfo {pages} {186802} (\bibinfo {year}
			{2018})}\BibitemShut {NoStop}%
		\bibitem [{\citenamefont {Xu}\ \emph {et~al.}(2013)\citenamefont {Xu},
			\citenamefont {Yan}, \citenamefont {Zhang}, \citenamefont {Wang},
			\citenamefont {Xu}, \citenamefont {Tang}, \citenamefont {Duan},\ and\
			\citenamefont {Zhang}}]{PhysRevLett.111.136804}%
		\BibitemOpen
		\bibfield  {author} {\bibinfo {author} {\bibfnamefont {Y.}~\bibnamefont
				{Xu}}, \bibinfo {author} {\bibfnamefont {B.}~\bibnamefont {Yan}}, \bibinfo
			{author} {\bibfnamefont {H.-J.}\ \bibnamefont {Zhang}}, \bibinfo {author}
			{\bibfnamefont {J.}~\bibnamefont {Wang}}, \bibinfo {author} {\bibfnamefont
				{G.}~\bibnamefont {Xu}}, \bibinfo {author} {\bibfnamefont {P.}~\bibnamefont
				{Tang}}, \bibinfo {author} {\bibfnamefont {W.}~\bibnamefont {Duan}}, \ and\
			\bibinfo {author} {\bibfnamefont {S.-C.}\ \bibnamefont {Zhang}},\ }\href
		{\doibase 10.1103/PhysRevLett.111.136804} {\bibfield  {journal} {\bibinfo
				{journal} {Phys. Rev. Lett.}\ }\textbf {\bibinfo {volume} {111}},\ \bibinfo
			{pages} {136804} (\bibinfo {year} {2013})}\BibitemShut {NoStop}%
		\bibitem [{\citenamefont {Rohlfing}\ and\ \citenamefont
			{Louie}(2000)}]{PhysRevB.62.4927}%
		\BibitemOpen
		\bibfield  {author} {\bibinfo {author} {\bibfnamefont {M.}~\bibnamefont
				{Rohlfing}}\ and\ \bibinfo {author} {\bibfnamefont {S.~G.}\ \bibnamefont
				{Louie}},\ }\href {\doibase 10.1103/PhysRevB.62.4927} {\bibfield  {journal}
			{\bibinfo  {journal} {Phys. Rev. B}\ }\textbf {\bibinfo {volume} {62}},\
			\bibinfo {pages} {4927} (\bibinfo {year} {2000})}\BibitemShut {NoStop}%
		\bibitem [{\citenamefont {Scharf}\ \emph {et~al.}(2017)\citenamefont {Scharf},
			\citenamefont {Xu}, \citenamefont {Matos-Abiague},\ and\ \citenamefont
			{\ifmmode \check{Z}\else \v{Z}\fi{}uti\ifmmode~\acute{c}\else
				\'{c}\fi{}}}]{PhysRevLett.119.127403}%
		\BibitemOpen
		\bibfield  {author} {\bibinfo {author} {\bibfnamefont {B.}~\bibnamefont
				{Scharf}}, \bibinfo {author} {\bibfnamefont {G.}~\bibnamefont {Xu}}, \bibinfo
			{author} {\bibfnamefont {A.}~\bibnamefont {Matos-Abiague}}, \ and\ \bibinfo
			{author} {\bibfnamefont {I.}~\bibnamefont {\ifmmode \check{Z}\else
					\v{Z}\fi{}uti\ifmmode~\acute{c}\else \'{c}\fi{}}},\ }\href {\doibase
			10.1103/PhysRevLett.119.127403} {\bibfield  {journal} {\bibinfo  {journal}
				{Phys. Rev. Lett.}\ }\textbf {\bibinfo {volume} {119}},\ \bibinfo {pages}
			{127403} (\bibinfo {year} {2017})}\BibitemShut {NoStop}%
		\bibitem [{\citenamefont {Cao}\ \emph {et~al.}(2018)\citenamefont {Cao},
			\citenamefont {Wu},\ and\ \citenamefont {Louie}}]{PhysRevLett.120.087402}%
		\BibitemOpen
		\bibfield  {author} {\bibinfo {author} {\bibfnamefont {T.}~\bibnamefont
				{Cao}}, \bibinfo {author} {\bibfnamefont {M.}~\bibnamefont {Wu}}, \ and\
			\bibinfo {author} {\bibfnamefont {S.~G.}\ \bibnamefont {Louie}},\ }\href
		{\doibase 10.1103/PhysRevLett.120.087402} {\bibfield  {journal} {\bibinfo
				{journal} {Phys. Rev. Lett.}\ }\textbf {\bibinfo {volume} {120}},\ \bibinfo
			{pages} {087402} (\bibinfo {year} {2018})}\BibitemShut {NoStop}%
		\bibitem [{\citenamefont {Zhang}\ \emph {et~al.}(2018)\citenamefont {Zhang},
			\citenamefont {Shan},\ and\ \citenamefont {Xiao}}]{PhysRevLett.120.077401}%
		\BibitemOpen
		\bibfield  {author} {\bibinfo {author} {\bibfnamefont {X.}~\bibnamefont
				{Zhang}}, \bibinfo {author} {\bibfnamefont {W.-Y.}\ \bibnamefont {Shan}}, \
			and\ \bibinfo {author} {\bibfnamefont {D.}~\bibnamefont {Xiao}},\ }\href
		{\doibase 10.1103/PhysRevLett.120.077401} {\bibfield  {journal} {\bibinfo
				{journal} {Phys. Rev. Lett.}\ }\textbf {\bibinfo {volume} {120}},\ \bibinfo
			{pages} {077401} (\bibinfo {year} {2018})}\BibitemShut {NoStop}%
		\bibitem [{\citenamefont {Zhu}\ \emph {et~al.}(2019)\citenamefont {Zhu},
			\citenamefont {Tu}, \citenamefont {Tong},\ and\ \citenamefont
			{Yao}}]{doi:10.1126/sciadv.aau6120}%
		\BibitemOpen
		\bibfield  {author} {\bibinfo {author} {\bibfnamefont {Q.}~\bibnamefont
				{Zhu}}, \bibinfo {author} {\bibfnamefont {M.~W.-Y.}\ \bibnamefont {Tu}},
			\bibinfo {author} {\bibfnamefont {Q.}~\bibnamefont {Tong}}, \ and\ \bibinfo
			{author} {\bibfnamefont {W.}~\bibnamefont {Yao}},\ }\href {\doibase
			10.1126/sciadv.aau6120} {\bibfield  {journal} {\bibinfo  {journal} {Sci.
					Adv.}\ }\textbf {\bibinfo {volume} {5}},\ \bibinfo {pages} {eaau6120}
			(\bibinfo {year} {2019})}\BibitemShut {NoStop}%
		\bibitem [{\citenamefont {Yu}\ \emph {et~al.}(2011)\citenamefont {Yu},
			\citenamefont {Qi}, \citenamefont {Bernevig}, \citenamefont {Fang},\ and\
			\citenamefont {Dai}}]{PhysRevB.84.075119}%
		\BibitemOpen
		\bibfield  {author} {\bibinfo {author} {\bibfnamefont {R.}~\bibnamefont
				{Yu}}, \bibinfo {author} {\bibfnamefont {X.~L.}\ \bibnamefont {Qi}}, \bibinfo
			{author} {\bibfnamefont {A.}~\bibnamefont {Bernevig}}, \bibinfo {author}
			{\bibfnamefont {Z.}~\bibnamefont {Fang}}, \ and\ \bibinfo {author}
			{\bibfnamefont {X.}~\bibnamefont {Dai}},\ }\href {\doibase
			10.1103/PhysRevB.84.075119} {\bibfield  {journal} {\bibinfo  {journal} {Phys.
					Rev. B}\ }\textbf {\bibinfo {volume} {84}},\ \bibinfo {pages} {075119}
			(\bibinfo {year} {2011})}\BibitemShut {NoStop}%
		\bibitem [{\citenamefont {Weng}\ \emph {et~al.}(2015)\citenamefont {Weng},
			\citenamefont {Yu}, \citenamefont {Hu}, \citenamefont {Dai},\ and\
			\citenamefont {Fang}}]{doi:10.1080/00018732.2015.1068524}%
		\BibitemOpen
		\bibfield  {author} {\bibinfo {author} {\bibfnamefont {H.}~\bibnamefont
				{Weng}}, \bibinfo {author} {\bibfnamefont {R.}~\bibnamefont {Yu}}, \bibinfo
			{author} {\bibfnamefont {X.}~\bibnamefont {Hu}}, \bibinfo {author}
			{\bibfnamefont {X.}~\bibnamefont {Dai}}, \ and\ \bibinfo {author}
			{\bibfnamefont {Z.}~\bibnamefont {Fang}},\ }\href {\doibase
			10.1080/00018732.2015.1068524} {\bibfield  {journal} {\bibinfo  {journal}
				{Adv. Phys.}\ }\textbf {\bibinfo {volume} {64}},\ \bibinfo {pages} {227}
			(\bibinfo {year} {2015})}\BibitemShut {NoStop}%
		\bibitem [{\citenamefont {Wang}\ \emph
			{et~al.}(2019{\natexlab{b}})\citenamefont {Wang}, \citenamefont {O.},
			\citenamefont {Wang},\ and\ \citenamefont {Xing}}]{WOS:000455595400033}%
		\BibitemOpen
		\bibfield  {author} {\bibinfo {author} {\bibfnamefont {R.}~\bibnamefont
				{Wang}}, \bibinfo {author} {\bibfnamefont {E.}~\bibnamefont {O.}}, \bibinfo
			{author} {\bibfnamefont {B.}~\bibnamefont {Wang}}, \ and\ \bibinfo {author}
			{\bibfnamefont {D.~Y.}\ \bibnamefont {Xing}},\ }\href {\doibase
			10.1038/s41467-018-08203-9} {\bibfield  {journal} {\bibinfo  {journal} {Nat.
					Commun.}\ }\textbf {\bibinfo {volume} {10}} (\bibinfo {year}
			{2019}{\natexlab{b}}),\ 10.1038/s41467-018-08203-9}\BibitemShut {NoStop}%
		\bibitem [{\citenamefont {Alicea}(2012)}]{WOS:000305907700002}%
		\BibitemOpen
		\bibfield  {author} {\bibinfo {author} {\bibfnamefont {J.}~\bibnamefont
				{Alicea}},\ }\href {\doibase 10.1088/0034-4885/75/7/076501} {\bibfield
			{journal} {\bibinfo  {journal} {Rep. Prog. Phys.}\ }\textbf {\bibinfo
				{volume} {75}} (\bibinfo {year} {2012}),\
			10.1088/0034-4885/75/7/076501}\BibitemShut {NoStop}%
		\bibitem [{\citenamefont {Fu}\ and\ \citenamefont
			{Kane}(2008)}]{PhysRevLett.100.096407}%
		\BibitemOpen
		\bibfield  {author} {\bibinfo {author} {\bibfnamefont {L.}~\bibnamefont
				{Fu}}\ and\ \bibinfo {author} {\bibfnamefont {C.~L.}\ \bibnamefont {Kane}},\
		}\href {\doibase 10.1103/PhysRevLett.100.096407} {\bibfield  {journal}
			{\bibinfo  {journal} {Phys. Rev. Lett.}\ }\textbf {\bibinfo {volume} {100}},\
			\bibinfo {pages} {096407} (\bibinfo {year} {2008})}\BibitemShut {NoStop}%
		\bibitem [{\citenamefont {Seradjeh}(2012)}]{PhysRevB.86.121101}%
		\BibitemOpen
		\bibfield  {author} {\bibinfo {author} {\bibfnamefont {B.}~\bibnamefont
				{Seradjeh}},\ }\href {\doibase 10.1103/PhysRevB.86.121101} {\bibfield
			{journal} {\bibinfo  {journal} {Phys. Rev. B}\ }\textbf {\bibinfo {volume}
				{86}},\ \bibinfo {pages} {121101} (\bibinfo {year} {2012})}\BibitemShut
		{NoStop}%
		\bibitem [{\citenamefont {Murakami}\ \emph {et~al.}(2020)\citenamefont
			{Murakami}, \citenamefont {Gole{\v{z}}}, \citenamefont {Kaneko},
			\citenamefont {Koga}, \citenamefont {Millis},\ and\ \citenamefont
			{Werner}}]{murakami2020collective}%
		\BibitemOpen
		\bibfield  {author} {\bibinfo {author} {\bibfnamefont {Y.}~\bibnamefont
				{Murakami}}, \bibinfo {author} {\bibfnamefont {D.}~\bibnamefont
				{Gole{\v{z}}}}, \bibinfo {author} {\bibfnamefont {T.}~\bibnamefont {Kaneko}},
			\bibinfo {author} {\bibfnamefont {A.}~\bibnamefont {Koga}}, \bibinfo {author}
			{\bibfnamefont {A.~J.}\ \bibnamefont {Millis}}, \ and\ \bibinfo {author}
			{\bibfnamefont {P.}~\bibnamefont {Werner}},\ }\href@noop {} {\bibfield
			{journal} {\bibinfo  {journal} {Phys. Rev. B}\ }\textbf {\bibinfo {volume}
				{101}},\ \bibinfo {pages} {195118} (\bibinfo {year} {2020})}\BibitemShut
		{NoStop}%
		\bibitem [{\citenamefont {Gole{\v{z}}}\ \emph {et~al.}(2020)\citenamefont
			{Gole{\v{z}}}, \citenamefont {Sun}, \citenamefont {Murakami}, \citenamefont
			{Georges},\ and\ \citenamefont {Millis}}]{golevz2020nonlinear}%
		\BibitemOpen
		\bibfield  {author} {\bibinfo {author} {\bibfnamefont {D.}~\bibnamefont
				{Gole{\v{z}}}}, \bibinfo {author} {\bibfnamefont {Z.}~\bibnamefont {Sun}},
			\bibinfo {author} {\bibfnamefont {Y.}~\bibnamefont {Murakami}}, \bibinfo
			{author} {\bibfnamefont {A.}~\bibnamefont {Georges}}, \ and\ \bibinfo
			{author} {\bibfnamefont {A.~J.}\ \bibnamefont {Millis}},\ }\href@noop {}
		{\bibfield  {journal} {\bibinfo  {journal} {Phys. Rev. Lett.}\ }\textbf
			{\bibinfo {volume} {125}},\ \bibinfo {pages} {257601} (\bibinfo {year}
			{2020})}\BibitemShut {NoStop}%
		\bibitem [{\citenamefont {Tanabe}\ \emph {et~al.}(2021)\citenamefont {Tanabe},
			\citenamefont {Kaneko},\ and\ \citenamefont {Ohta}}]{tanabe2021third}%
		\BibitemOpen
		\bibfield  {author} {\bibinfo {author} {\bibfnamefont {T.}~\bibnamefont
				{Tanabe}}, \bibinfo {author} {\bibfnamefont {T.}~\bibnamefont {Kaneko}}, \
			and\ \bibinfo {author} {\bibfnamefont {Y.}~\bibnamefont {Ohta}},\ }\href@noop
		{} {\bibfield  {journal} {\bibinfo  {journal} {Phys. Rev. B}\ }\textbf
			{\bibinfo {volume} {104}},\ \bibinfo {pages} {245103} (\bibinfo {year}
			{2021})}\BibitemShut {NoStop}%
		\bibitem [{\citenamefont {Kaneko}\ \emph {et~al.}(2021)\citenamefont {Kaneko},
			\citenamefont {Sun}, \citenamefont {Murakami}, \citenamefont {Gole{\v{z}}},\
			and\ \citenamefont {Millis}}]{kaneko2021bulk}%
		\BibitemOpen
		\bibfield  {author} {\bibinfo {author} {\bibfnamefont {T.}~\bibnamefont
				{Kaneko}}, \bibinfo {author} {\bibfnamefont {Z.}~\bibnamefont {Sun}},
			\bibinfo {author} {\bibfnamefont {Y.}~\bibnamefont {Murakami}}, \bibinfo
			{author} {\bibfnamefont {D.}~\bibnamefont {Gole{\v{z}}}}, \ and\ \bibinfo
			{author} {\bibfnamefont {A.~J.}\ \bibnamefont {Millis}},\ }\href@noop {}
		{\bibfield  {journal} {\bibinfo  {journal} {Phys. Rev. Lett.}\ }\textbf
			{\bibinfo {volume} {127}},\ \bibinfo {pages} {127402} (\bibinfo {year}
			{2021})}\BibitemShut {NoStop}%
		\bibitem [{\citenamefont {Dong}\ and\ \citenamefont
			{Li}(2023)}]{dong2023robust}%
		\BibitemOpen
		\bibfield  {author} {\bibinfo {author} {\bibfnamefont {S.}~\bibnamefont
				{Dong}}\ and\ \bibinfo {author} {\bibfnamefont {Y.}~\bibnamefont {Li}},\
		}\href@noop {} {\bibfield  {journal} {\bibinfo  {journal} {arXiv}\ }\textbf
			{\bibinfo {volume} {2303.16627}} (\bibinfo {year} {2023})}\BibitemShut
		{NoStop}%
		\bibitem [{\citenamefont {Giannozzi}\ \emph {et~al.}(2009)\citenamefont
			{Giannozzi} \emph {et~al.}}]{giannozzi2009quantum}%
		\BibitemOpen
		\bibfield  {author} {\bibinfo {author} {\bibfnamefont {P.}~\bibnamefont
				{Giannozzi}} \emph {et~al.},\ }\href@noop {} {\bibfield  {journal} {\bibinfo
				{journal} {J. Phys. Condens. Matter}\ }\textbf {\bibinfo {volume} {21}},\
			\bibinfo {pages} {395502} (\bibinfo {year} {2009})}\BibitemShut {NoStop}%
		\bibitem [{\citenamefont {Sangalli}\ \emph {et~al.}(2019)\citenamefont
			{Sangalli} \emph {et~al.}}]{sangalli2019many}%
		\BibitemOpen
		\bibfield  {author} {\bibinfo {author} {\bibfnamefont {D.}~\bibnamefont
				{Sangalli}} \emph {et~al.},\ }\href@noop {} {\bibfield  {journal} {\bibinfo
				{journal} {J. Phys. Condens. Matter}\ }\textbf {\bibinfo {volume} {31}},\
			\bibinfo {pages} {325902} (\bibinfo {year} {2019})}\BibitemShut {NoStop}%
	\end{thebibliography}
	%

\end{document}